\documentclass[aps,prb,twocolumn]{revtex4}

\usepackage{graphicx}

\begin{document}


\title{Statistics of low-energy levels of a one-dimensional
weakly localized Frenkel exciton: A numerical study}

\author{Andrei V. Malyshev}
\address{Ioffe Physiko-Technical Institute, 26 Politechnicheskaya str.,
194021 Saint-Petersburg, Russia}

\author{Victor A. Malyshev}
\address{National Research Center "Vavilov State Optical Institute",
Birzhevaya Liniya 12, 199034 Saint-Petersburg, Russia}

\date{\today}


\begin{abstract}

Numerical study of the one-dimensional Frenkel Hamiltonian with
on-site randomness is carried out. We focus on the statistics of
the energy levels near the lower exciton band edge, i. e. those
determining optical response. We found that the distribution of
the energy spacing between the states that are well localized at
the same segment is characterized by non-zero mean, i.e. these
states undergo repulsion. This repulsion results in a local
discrete energy structure of a localized Frenkel exciton. On the
contrary, the energy spacing distribution for weakly overlapping
local ground states (the states with no nodes within their
localization segments) that are localized at different segments
has zero mean and shows almost no repulsion. The typical width of
the latter distribution is of the same order as the typical
spacing in the local discrete energy structure, so that this local
structure is hidden; it does not reveal itself neither in the
density of states nor in the linear absorption spectra. However,
this structure affects the two-exciton transitions involving the
states of the same segment and can be observed by the pump-probe
spectroscopy. We analyze also the disorder degree scaling of the
first and second momenta of the distributions.

\end{abstract}


\maketitle

\section{Introduction}
\label{intro}

Since the pioneering work by Mott and Twose~\cite{Mott61} it is
well known that all states in one dimension (1D) are localized in
the presence of uncorrelated
disorder,~\cite{Flores1,Dunlap,Bellani99} which means that a
quasi-particle wave function has a finite amplitude within a
finite space interval and vanishes outside. The size of this
interval (localization length) increases with the decrease of
disorder. The 1D localization theorem was supported later by the
one-parameter scaling theory of Abrahams {\it et
al.}~\cite{Abrahams79} (see a comprehensive review in
Ref.~\onlinecite{Kramer93}).

The concept of 1D localization have been successfully applied to
explanation of the optical properties of linear molecular
aggregates and conjugated polymers, in which the elementary
excitations are Frenkel excitons (for a comprehensive review see
Refs.~\onlinecite{Fidder93a,Spano94,Knoester96}). The one-exciton
states that are closer to the bottom of the exciton energy band
couple better to the light and hence determine the linear optical
response. For this reason the energy (and wave function)
statistics of the lower states has been attracting great deal of
attention. In Ref.~\onlinecite{Malyshev91} the heuristic concept of
the hidden low-energy structure of a 1D localized exciton was put
forward. This concept was partly supported by the numerical
simulations.~\cite{Malyshev95,Shimizu98,Malyshev99,Bakalis99}
However, it has not been proved directly by detailed statistical study.

The basic idea of the hidden low-energy structure concept is as
follows. The lower energy one-exciton eigenfunctions obtained for
a fixed realization of disorder can be grouped in sets of two (or
sometimes more) states which are localized at the same segment of
the linear chain. These segments are almost non-overlapping. The
typical length of a segment $N^{*}$ depends on the degree of
disorder (the quantity $N^*$ is often called the number of
coherently bound molecules~\cite{Knapp84}). The energies of the
states belonging to the same set are well separated, so that they
form a local discrete energy structure. The mean energy of such
local manifolds is distributed within the energy interval of the
same order as the typical energy spacing between the levels of a
local manifold. For this reason the local energy structure appears
to be hidden; it does not manifest itself neither in the density
of states nor in the linear absorption spectra. Nevertheless, it
can be revealed from the optical response involving two-exciton
transitions.~\cite{Malyshev95,Minoshima94,Bakalis99}

In this paper we primarily focus on the direct proof of the heuristic ideas
put forward in Refs.~\onlinecite{Malyshev91,Malyshev95}. To do this we study
in detail the low-energy levels statistics for a weakly localized 1D Frenkel
exciton. The bulk of the paper is organized as follows. In the next Section
the model Hamiltonian is described. In Sec.~\ref{Numerics} the numerical
results of the low-energy level statistics simulation are discussed.
Section~\ref{Sum} summarizes the paper and comments on the importance of the
local discrete energy structure for optical response from the disordered
Frenkel exciton systems.

\section{Model Hamiltonian}
\label{Hamiltonian}

We consider ${\cal N}$ (${\cal N}\gg 1$) optically active two-level
molecules forming a regular 1D lattice with unity spacing. The corresponding
Frenkel exciton Hamiltonian reads:~\cite{Davydov71}
\begin{equation}
    H = \sum_{i=1}^{\cal N}\> E_i |i \rangle \langle i| +
    \sum_{i,j=1}^{\cal N}\>J_{ij}^{} |i \rangle \langle j|\ .
    \label{H}
\end{equation}
Here $|i \rangle$ is the state vector of the excited molecule with the
energy $E_i$ ($1\le i,j \le {\cal N}$). The energy is assumed to be the
Gaussian uncorrelated (for different sites) stochastic variable with
zero mean and the
standard deviation $\Delta$. On the contrary, the hopping integrals
$J_{ij}$\ ($J_{ii}\equiv 0$) are considered to be non-fluctuating. These
integrals are of the dipole-dipole origin: $J_{ij}=J/|i-j|^{3}$, where $J$
is the nearest-neighbor coupling. The quantity $\Delta/|J|$ will be referred
to as the degree of disorder. Hereafter, we assume $J$ to be negative, which
corresponds to the case of the
J-aggregates.~\cite{Fidder93a,Spano94,Knoester96} In this case the states
coupled to the light are those at the bottom of the exciton band (see, for
instance, Ref.~\onlinecite{Spano94}).

As it was pointed out in Ref.~\onlinecite{Fidder91} the long-range
dipole-dipole terms strongly affect the exciton-radiative-rate enhancement
factor in the presence of disorder. In this paper we discuss both the
nearest-neighbor (NN) approximation and the effects of the long-range
dipole-dipole (DD) coupling.

Let us briefly remind the reader the most important features of the
one-exciton energy spectrum and eigenfunctions in the absence of disorder
$(\Delta=0)$. They can be found as the solutions of the following
eigenvalue problem:
\begin{mathletters}
    \label{1}
    \begin{equation}
        |K\rangle = \sum_{i=1}^{\cal N} \varphi_{Ki}|i\rangle\ ,
    \label{k}
    \end{equation}
    \begin{equation}
        \sum_{i=1}^{\cal N} H^{ij}\varphi_{Kj} = E_K \varphi_{Ki} \ ,
        \label{EigFunc}
    \end{equation}
\end{mathletters}

\noindent where $H^{ij} = \langle i| H |j\rangle$, $\varphi_{K}$ and $E_K$
are the eigenfunction and eigenenergy of the one-exciton state
$|K\rangle$\ respectively; the quantity $K = \pi k/ ({\cal N} + 1)$,
where $k \in [1, {\cal N}]$, plays the role of the exciton wavenumber.

The eigenenergies $E_K$ for the exact dipole-dipole model
for ${\cal N} \gg 1$ can be written as:~\cite{Malyshev95}
\begin{mathletters}
    \label{2}
    \begin{equation}
        E_K=-2|J| \sum_{i=1}^{\cal N}{\cos\left(Ki\right) \over i^3}
        + {\cal O}({\cal N}^{-1})\ .
       \label{Ek}
    \end{equation}
The energy spectrum in the NN model can be obtained from
Eq.~\ref{Ek} by keeping only the $i \equiv 1$ term and appears to
be very different from the exact DD spectrum (see below). In
contrast to this, the eigenfunctions $\varphi_{K}$ are almost the
same in both models:~\cite{Fidder91}
    \begin{equation}
        \varphi_{Ki} = \left( {2\over {\cal N}+1}\right)^{1/2}
        \sin\left(Ki\right)\ .
        \label{phi}
    \end{equation}
\end{mathletters}

As it was mentioned above we are especially interested in the
exciton spectrum at the bottom of the exciton band, i. e. for $K
\ll 1$. In this limit the spectrum was obtained and discussed in
Ref.~\onlinecite{Malyshev95}:
\begin{equation}
    E_K=-2|J|\zeta (3)+|J|\Biggl({3\over 2}-\ln K\Biggr)K^2 , \quad K\ll 1\ ,
    \label{BottomLR}
\end{equation}
where $\zeta (3)=\sum_{i=1}^\infty i^{-3}\approx 1.202$. The
corresponding expression in the NN approximation (the term with $i = 1$
in the sum in Eq.~(\ref{Ek})) reads:
\begin{equation}
    E_K=-2|J| - |J| K^2\ .
    \label{BottomNN}
\end{equation}

Straightforward comparison of the two equations shows that the
long-range DD terms affect the exciton low-energy spectrum
considerably. First, the exciton band bottom is red-shifted in the
DD model with respect to the NN case: from $-2|J|$ to
approximately $-2.404 |J|$. Second, the $K$-dependence of the
energy is stronger in the DD case due to the logarithmic factor
in~(\ref{BottomLR}). In particular, the energy difference between
the two lowest exciton states (with $K_2 = 2\pi/ ({\cal N} + 1)$
and $K_1 = \pi/ ({\cal N} + 1)$) is greater when the long-range
terms are considered.

As well as the eigenfunctions $\varphi_{K}$, the oscillator strengths are
almost the same in both models. For the transitions from the ground to lower
exciton states they can be written as (see, for instance,
Ref.~\onlinecite{Spano94})
\begin{equation}
    f_{K} =
    \left(\sum_{i=1}^{\cal N} \varphi_{Ki}\right)^2
    = {1 -(-1)^{k} \over 2}\, {8\over K^2}\ ,
    \quad k = 1,2,...
\label{fk}
\end{equation}
Here the dipole momentum of the optical transition is taken to be unity.
Thus, the oscillator strength of the transition from the ground state to the
lowest exciton state ($K = K_1 = \pi/ ({\cal N} + 1)$),
\begin{equation}
    f_{K_1} = {8\over \pi^2} ({\cal N} + 1)
    \approx 0.81 ({\cal N} + 1),
\label{fk1}
\end{equation}
is proportional to the number of sites in the chain ${\cal N}$ and carries
81\% of the total oscillator strength (which is equal to ${\cal N}$).

\section{Numerical results and discussion}
\label{Numerics}

In this section statistical properties of the one-exciton low-energy
eigenfunctions and eigenenergies of the Hamiltonian (\ref{H}) are analyzed
numerically. We focus on the distributions of the localization segment
length and energy spacings between lower states. The disorder degree scaling
of the first momenta of these distributions will also be the subject of our
interest. This requires numerical solution of the eigenvalue problem
(\ref{1}) for a large number of disorder realizations.

\begin{figure}[h!]
\includegraphics[height=7in]{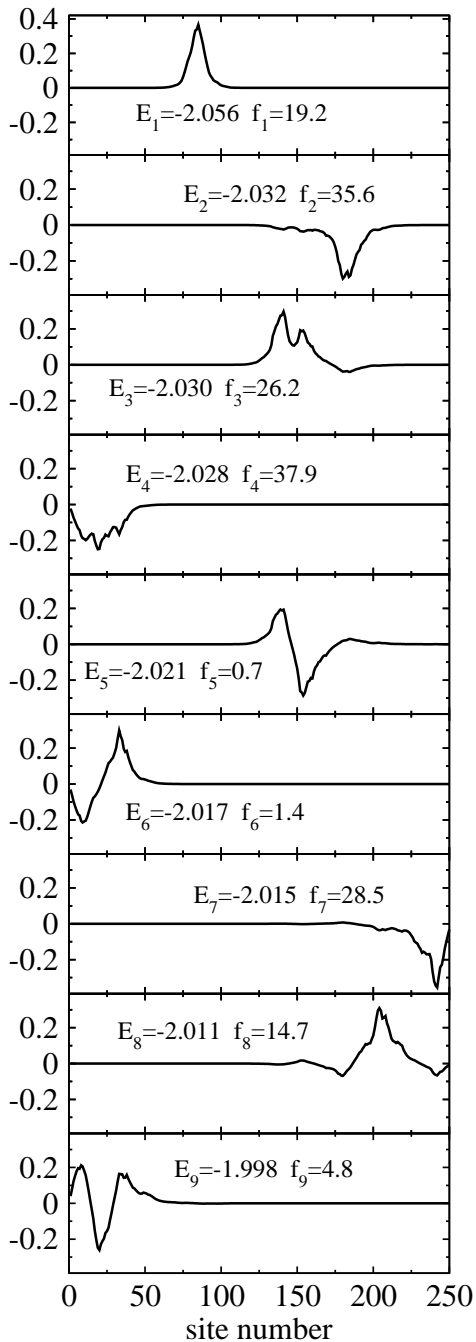}
\caption{
The typical realization of the low-energy eigenfunctions of the
Hamiltonian~(\protect\ref{H}) for the 250 sites linear and the disorder
degree $\Delta/|J|=0.1$. Each plot also shows the corresponding energy of
the state $E_\nu$ (in units of $|J|$) and its oscillator strength $f_\nu$.
}
\label{fig1}
\end{figure}

\subsection{Heuristic arguments}
\label{Arguments}

Before doing the detailed statistical analyses let us look at a
typical realization of the eigenstates (for a fixed stochastic
distribution of site energies). Figure~\ref{fig1} presents an
example showing the nine lowest eigenfunctions $\varphi_{\nu}$
(these are real in our case) and the corresponding eigenenergies
$E_\nu$. This set of wave functions was obtained for 250 sites
chain in the NN model for the disorder degree $\Delta/|J|=0.1$.
Each plot also displays the corresponding oscillator strength
$f_\nu = \left(\sum_{i=1}^{\cal N} \varphi_{\nu i}\right)^2.$

Several typical features of the lower eigenstates can clearly be
seen from Fig.~\ref{fig1}. First, all eigenfunctions are well
localized at different segments of the chain. The lengths of these
segments $N$ are smaller than the total chain length ${\cal N}$.
Some of the eigenfunctions are localized at the same segment ({\it
e. g.} manifold of states 4, 6 and 9) and weakly overlap with the
other states. Second, one of the states of each particular (local)
manifold has no nodes within its localization segment and hence
have high oscillator strength ({\it e. g.} state 4) while the
other wave functions from the manifold ({\it e. g.} states 6 and
9) have nodes within the segment and low oscillator strengths.
This allows to classify such states as local ground and local
excited states respectively. These states are similar to the
states of a regular non-disordered chain (see Eq.~(\ref{phi}) at
$k=1$ and $k=2,\,3$). Finally, it is important to note that a
local ground state can have higher energy than a local excited
state which is localized at a different segment. For example, the
local ground state $7$ has higher energy ($E_7=-2.015|J|$) than
the local excited states $6$ ($E_6=-2.021|J|$) and $5$
($E_5=-2.017|J|$). This raises the question about the correlations
between the local ground and excited states from one manifold, on
the one hand, and between the states from different manifolds, on
the other hand.

It is clear from these arguments that grouping the states together into
local manifolds is a principal problem; correct criteria are needed to
select the states which are localized at the chain segments. When such
grouping is done the statistical analysis we are interested in becomes
simple.

\subsection{Selection criteria}
\label{criteria}

To select the states which are localized at the same segment first we need
to find the local ground states. Such states $\varphi_{\nu}$ satisfy the
inequality:
\begin{equation}
    \left|\sum_i \varphi_{\nu i}\,|\varphi_{\nu i}|\right| \ge C_0 \ .
    \label{gs_criterion}
\end{equation}
For the constant $C_0$ we use the value $C_0=0.95$, seeking, in other words,
for the states which contain at least 95\% of the wave function density in
the main peak. The states $1-4$ and $7$ in the Fig.~\ref{fig1} meet this
criterion and can be considered as the local ground states.

Then we find all local excited states. In order to do this we proceed as
follows: for each local ground state $\nu$ ($\nu = 1-4,7$ in our example) we
seek for the states $\nu^\prime \neq \nu$ which are localized at the same
segment as the state $\nu$. These (local excited) states should
simultaneously meet two criteria. The first is similar to the criterion
given by Eq.~(\ref{gs_criterion}):
\begin{equation}
    \left|\sum_i \varphi_{\nu i}\,|\varphi_{\nu^\prime i}|\right|\ge
    C_1 \ ,
    \label{es_criterion1}
\end{equation}
where $i$ runs over all sites of the chain. The value for the cut-off
constant $C_1$ will be discussed later. The second criterion is to be
applied to the Fourier transforms $\phi_{\nu}$ corresponding to the wave
functions $\varphi_{\nu}$:
\begin{equation}
    \sum_q |\phi_{\nu q}|\,|\phi_{\nu^\prime q}| \ge C_2 \ .
    \label{es_criterion2}
\end{equation}
The later is necessary to drop rapidly oscillating states $\nu^\prime$,
envelope functions of which are localized within the same segment as the
considered ground state $\nu$. The energy of such states is in the vicinity
of the exciton band maximum. Their spatial extension is of the same order of
magnitude as the localization length of the low-energy states, so these
high-energy states often meet the criterion (\ref{es_criterion1}).

The reasonable values for the cut-off parameters $C_1$ and $C_2$
can be estimated by considering the states of a rectangular
quantum well. In the limiting case of an infinitely deep quantum
well the overlap integral for the ground state (given by
Eq.~(\ref{phi}) at $k=1$) and the module of the first excited
state (Eq.~(\ref{phi}) at $k=2$) is equal to $I_m = 8/3\pi \approx
0.85$. In the more general case of a finite rectangular well the
value of this integral depends on the well depth and width. It
varies from about $I_{m}$ (for a very deep excited state) to
zero. Small values of the integral occur for very extended excited
states that are just appeared in the well.

Consider the well for which the probability to find the particle
in the well equals $0.95$ for the ground state, then for the first
excited state this probability equals about $0.75$. The overlap
integral of the type (\ref{es_criterion1}) for these states is
about $0.8$. This value gives an estimate for the parameter $C_1$.
The overlap integral of the moduli of the Fourier transforms of
these states is about $0.56$, which gives an estimate for the
$C_2$. For our estimations we considered the limiting case of a
rectangular well potential which varies infinitely rapidly at the
well boundaries. In a smoother potential the values of the overlap
integrals can be smaller. To account for this we use more relaxed
cut-off criteria with $C_1=0.7$ and $C_2=0.5$. It is also worth
noting that straightforward calculations revealed only weak
quantitative dependence of the final results on the values of the
cut-off parameters (see Appendix).

Applying the criteria (\ref{es_criterion1}) and (\ref{es_criterion2}) to our
sample set of eigenfunctions we find the following pairs of the local ground
and first excited states:
$\{3, 5\}$ and $\{4, 6\}$. This finding looks reasonable (see
Fig~\ref{fig1}). The states $4$ and $9$ also meet the above criteria.
Fig.~\ref{fig1} shows that the state $9$ has two well defined nodes within
the segment of localization of the local ground state $4$. It therefore can
be referred to as the second local excited state. However, the second and
higher local excited states will not be the subject of the present study.
Such higher excited states are usually localized at more extended segments
and can not be associated with a particular local ground state.

\pagebreak

\subsection{Statistics of the local ground states}
\label{gs_statistics}

Figure~\ref{fig1} clearly shows that the spatial extension of a local ground
state (its localization length $N$) fluctuates from one state to another.
Different quantities can be used to characterize the extension of a wave
function: (i) the inverse participation ratio (IPR), defined as $\sum_i
\,|\varphi_{\nu i}|^4$,\, (ii) the mean-square displacement $\left[\sum_i
\,|\varphi_{\nu i}|^2 (i - \bar i)^2\right]^{1/2}$, where $\bar i = \sum_i
\,|\varphi_{\nu i}|^2\, i$,\, (iii) the inverse of the Lyapunov exponent
$\lambda_\nu = - \lim_{i\rightarrow \infty}\, (1/2i) \ln |\varphi_{\nu
i}|^2$ and others.~\cite{Kramer93} Although these quantities yield slightly
different localization lengths, they are of the order of the number of sites
over which the wave function is extended.

For the local ground state, an alternative quantity can be used as the
measure of its extension. According to Eq.~(\ref{fk1}), the oscillator
strength of the ground state of the ideal linear chain is about the same as
the chain length. In other words, $f_\nu$ also carries information about the
extension of the corresponding local ground state wavefunction
$\varphi_\nu$. Furthermore, the oscillator strength can be measured
experimentally, e.g. it can be determined from the exciton spontaneous decay
kinetics~\cite{Fidder93a,Fidder91}). For this reason we use the oscillator
strength as the measure of localization of the local ground states.

\begin{figure}[h!]
\includegraphics[height=4in]{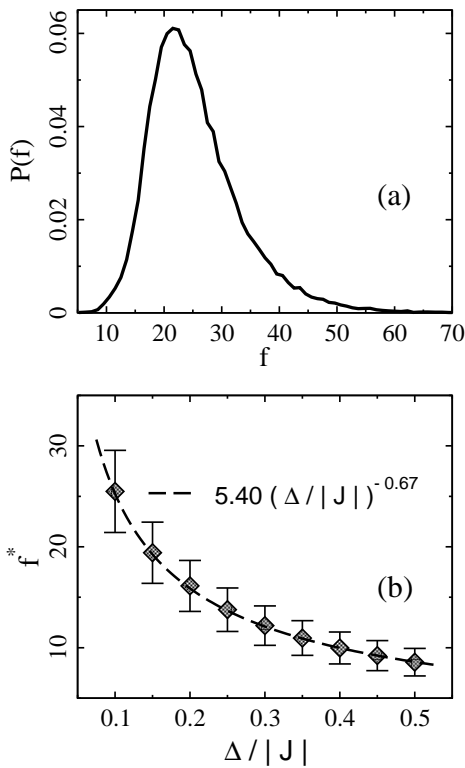}
\caption{
({\it a}) - The distribution $P(f)$ of the local ground state oscillator
strength $f$ obtained for a 400 sites chain and the disorder degree
$\Delta/|J|=0.1$ (5000 realizations of disorder was considered, each point
of the distribution is averaged over the interval (accumulation length)
$\delta = 1$).
({\it b}) - the disorder degree scaling of the first momentum $f^*
\equiv \int df P(f)f$ obtained for a 400 sites chain (300 realizations of
disorder was considered for each value of the disorder degree $\Delta/|J|$).
Diamonds present the numerical data while the dashed line gives the best
power-law fit. Error bars show the standard deviation of the distribution,
$\sigma\equiv [\int df P(f)(f - f^*)^2]^{1/2}$.
}
\label{fig2}
\end{figure}

The number of sites in the chain ${\cal N}$ should be much larger
than the typical oscillator strength $f^*$ (${\cal N} \gg f^*$).
As $f^*$ was found to be of the order of several tens for the
disorder degree $\Delta/|J|=0.1$ and decreases with increasing
$\Delta/|J|$, we consider $400$ cites chains (${\cal N} =400$).
Applying the selection criterion (\ref{gs_criterion}) we
calculated the $f$-distribution for the local ground states. The
result obtained in the NN approximation is presented in the
Fig.~\ref{fig2}. Figure \ref{fig2}({\it a}) shows the (typical)
oscillator strength distribution $P(f)$ obtained for a 400 sites
chain by averaging over 5000 realizations of disorder for the
disorder degree $\Delta/|J| = 0.1$. The main features of the
distribution are clearly seen in the figure: (i) - asymmetry with
respect to the expectation value of $f$, (ii) - long tail in the
large lengths region, and, on the contrary, (iii) - steep drop off
in the region of small values of $f$.

Figure ~\ref{fig2}({\it b}) presents the disorder degree scaling
of the first momentum of the $P(f)$ distribution, $f^* \equiv \int
df P(f)f$. Error bars in Fig.~\ref{fig2}({\it b}) show the
standard deviation $\sigma \equiv [\int df P(f)(f -
f^*)^2]^{1/2}$. The best fit (dashed line) to the numerical data
(diamonds) is given by the power-law approximation in the form
    \begin{equation}
        f^* \approx 5.40 \,\left({\Delta \over |J|}\right)^{-0.67}\ .
        \label{f*}
    \end{equation}

The theoretical estimate for the localization length was presented
in Refs.~\onlinecite{Malyshev91,Malyshev95} (see also
Ref.~\onlinecite{Bakalis00}) and was based on the following
arguments: An exciton localized at a segment of size $N^*$ feels a
reduced (exchange-narrowed) disorder $\sigma\equiv\Delta/\sqrt
{N^*}$;~\cite{Knapp84} this reduced disorder degree must be of the
same order of magnitude as the energy separation between the two
lowest local exciton levels, $E_2^* - E_1^* = 3\pi^2|J|/N^*$.
Indeed, if $E_2^* - E_1^*$ is smaller than $\sigma$ the two levels
would be strongly mixed by disorder and the corresponding exciton
wave functions would tend to reduce their extension. On the
contrary, if $E_2^* - E_1^*$ exceeds $\sigma$ then the disorder is
perturbative and the wave functions would tend to increase their
extension. Thus, the equality $E_2^* - E_1^* = \sigma$ gives the
self-consistent relationship between the spatial extension of the
exciton wave functions and the disorder degree which the localized
exciton feels. This relationship provides an estimate for the
localization length $N^*$:
    \begin{equation}
        N^* \approx \left({3\pi^2|J| \over \Delta}\right)^{2/3}
        \approx 9.57 \, \left({\Delta \over |J|}\right)^{-2/3} \ .
        \label{N*}
    \end{equation}
This theoretical scaling law is in good agreement with the scaling law of
$f^*$ obtained numerically, the latter being smaller by the factor of about
$2$. This is not surprising because, according to Eq.~(\ref{fk1}), the
oscillator strength is smaller than the localization length. The mean
localization length $N^*$ was also calculated ($N$ was defined as the number
of sites under the main peak of the wavefunction, for which $|\varphi_{\nu
i}| / max_i \{|\varphi_{\nu i}|\} > 0.1$). The mean oscillator strength
$f^*$ was found to be proportional to $N^*$: $f^* = c N^*$, where $c \approx
0.66$\ . Rewriting Eq.~(\ref{f*}) in terms of $N^*$ yields $N^*\approx 8.71 
\, (\Delta/|J|)^{-0.67}$, which is in excellent agreement with the 
theoretical estimate (\ref{N*}).

\subsection{Statistics of the energy levels near the lower exciton
band edge} \label{E_statistics}

\begin{figure}[h!]
\includegraphics[height=5.75in]{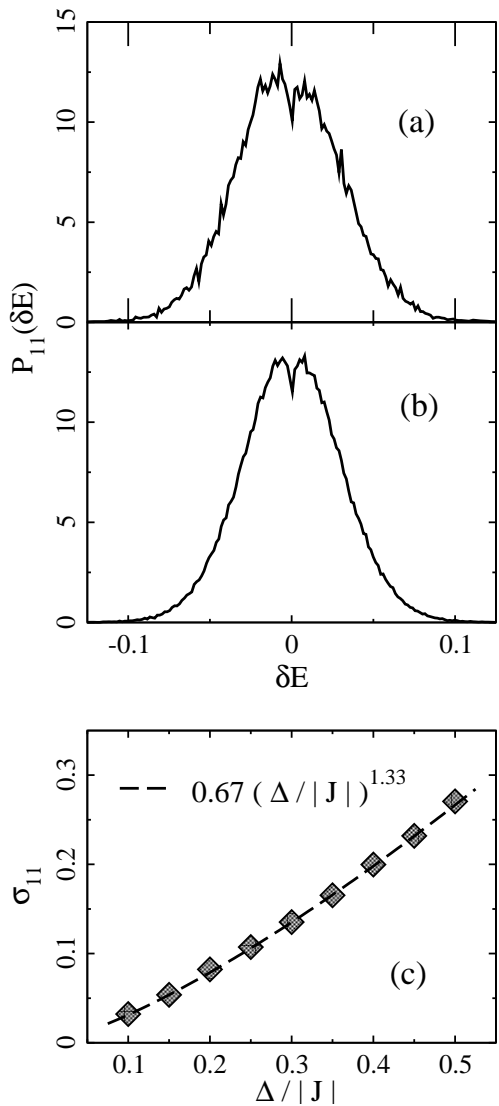}
\caption{
({\it a,b}) - The distributions of the energy spacing between the local
ground states obtained for a 400 sites chain and the disorder degree
$\Delta/|J|=0.1$ (5000 realizations of disorder was considered, each point
of the distributions is averaged over the interval (accumulation length)
$\delta = 0.0015$). The plot ({\it a}) shows the distribution 
$P_{11}(\delta E)$ for the adjacent local ground states (with the overlap 
integral $\sum_i\,|\varphi_{\nu i}|\,|\varphi_{\nu^\prime i}| \geq 0.01$). 
The distribution shown in the plot ({\it b}) was calculated for all local 
ground states.
({\it c}) - the disorder degree scaling of the standard deviation
$\sigma_{11} = [\int d(\delta E) (\delta E)^2 P_{11}(\delta E)]^{1/2}$ 
obtained for a 400 sites chain (300 realizations of
disorder was considered for each point). Diamonds present the
numerical data while the dashed line gives the best power-law fit.
All energies are in units of $|J|$.
}
\label{fig3}
\end{figure}

\begin{figure}[h!]
\includegraphics[height=4.5in]{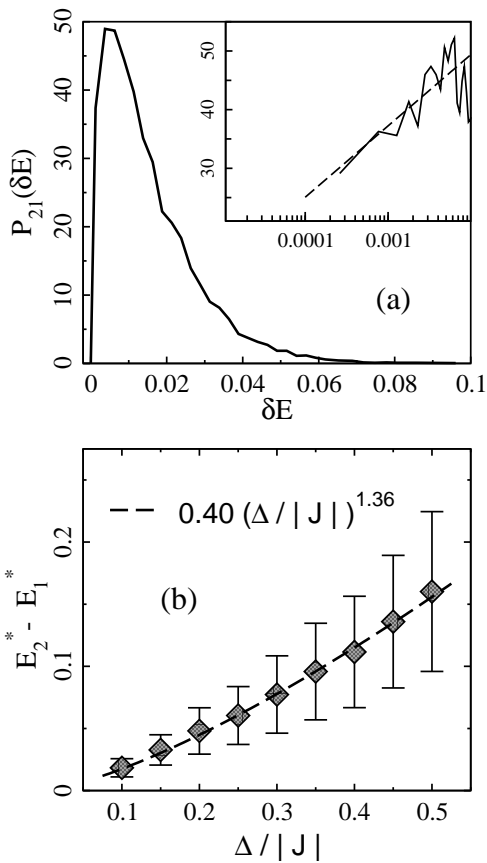}
\caption{
({\it a}) - The distribution of the energy spacing between the local ground
and first excited states of the same segment, $P_{21}(\delta E)$, obtained
for a 400 sites chain and the disorder degree $\Delta/|J|=0.1$ (10000
realizations of disorder is considered, each point of the distributions is
averaged over the interval (accumulation length) $\delta = 0.0025$). The
insert shows the behavior of the distribution function $P_{21}(\delta E)$ in
the region of very small energy spacings $\delta E$ (each point is averaged
over the interval $\delta = 0.0005$)
({\it b}) - The disorder degree scaling of the first momentum
$E^*_2 - E^*_1 \equiv \int d(\delta E) P_{21}(\delta E)\delta E$
obtained for a 400 sites chain (each point is averaged over 300
realizations of disorder). Diamonds present the numerical data
while the dashed line gives the best power-law fit. Error bars
show the standard deviation of the distribution,
$\sigma_{21} = [ \int d(\delta E) P_{21}(\delta E) (E_2^* - E_1^*
- \delta E)^2 ]^{1/2}$. All energies are in units of $|J|$.
}
\label{fig4}
\end{figure}

In this section we analyze two eigenenergy distributions which are
of primary interest for us. The first is the distribution of the
energy spacing between the local ground states of different
segments, while the second is the distribution of the energy
spacing between the local excited and ground states of the same
segment. We denote them as $P_{11}(\delta E)$ and $P_{21}(\delta
E)$ respectively. Utilizing the selection criteria of
Sec.~\ref{criteria}
(Eqs.~(\ref{gs_criterion})-(\ref{es_criterion2})) we calculated
these distributions. The results of these calculations are
presented in Figs.~\ref{fig3} and~\ref{fig4}.

Figures~\ref{fig3}({\it a}) and~\ref{fig3}({\it b}) present the (typical)
distribution of the energy spacing between the local ground states
$P_{11}(\delta E)$ obtained for a 400 sites chain by averaging over 5000
realization of disorder. The disorder degree $\Delta/|J|=0.1$\ and the NN
model was used. The plot ({\it a}) shows $P_{11}(\delta E)$ for the adjacent
(most overlapping) ground states only. These states were selected by the
criterion $\sum_i\,|\varphi_{\nu i}|\,|\varphi_{\nu^\prime i}| \geq 0.01$. The
distribution presented in Fig.~\ref{fig3}({\it b}), was obtained without the
above restriction, i. e. for all ground states.

The two distributions are very close to each other. The most
important feature of both is that they are nearly symmetric with respect to
zero energy spacing, in other words, they are characterized by zero
mean. The disorder degree scaling of the corresponding first momenta (not
presented here) confirms this fact. A weakly pronounced downfall in the
vicinity of the zero energy spacing of the distribution for the adjacent
states (plot ({\it a})) is due to the repulsion of close energy levels of
the overlapping states. Similarly, two degenerate states in a camel's back
potential are split into a doublet when interaction between them is
considered. This downfall tends to disappear when all ground state are
considered (see Fig.~\ref{fig3}({\it b})).

The plot ({\it c}) of Fig.~\ref{fig3} shows the disorder degree scaling of
the standard deviation of the distribution $P_{11}(\delta E)$, $\sigma_{11}
= [\int d(\delta E) (\delta E)^2 P_{11}(\delta E)]^{1/2}$, calculated for
all ground states. The best power-law fit (dashed line) to the numerical
data (diamonds) is given by
    \begin{equation}
        \sigma_{11}  \approx  0.67\;|J|\,\left(\Delta \over|J|\right)^{1.33}\ .
        \label{sigma}
    \end{equation}
The scaling law of $\sigma_{11}$ is close to the scaling law of
the one-exciton absorption
linewidth~\cite{Schreiber82,Koehler89,Boukahil90},  which supports
the idea that these quantities are proportional to each other.

\pagebreak

Figure~\ref{fig4}({\it a}) shows the (typical) distribution
$P_{21}(\delta E)$ of the energy spacing between the ground and
first excited states localized within the same segment. This
distribution was obtained for the same set of parameters (${\cal
N}=400$, $\Delta/|J|=0.1$, 10000 realizations). In contrast to the
$P_{11}(\delta E)$ (Fig.~\ref{fig3}({\it a, b})) the distribution
$P_{21}(\delta E)$ is asymmetric with respect to the expectation
value of $\delta E$. The high-energy-spacing shoulder of this distribution
is much longer than the shoulder in the region of small values of $\delta
E$. However, the most important features of the $P_{21}(\delta E)$ are the
following. First, it is characterized by non-zero mean, which unambiguously
shows that the energies of the pairs of states that are localized within the
same segment are highly correlated, in the sense that they undergo repulsion
resulting in the local discrete energy structure. Second, the distribution
$P_{21}(\delta E)$ drops down rapidly at $\delta E \simeq {\delta E}_{c}$
and vanishes below the energy ${\delta E}_{c}$. For the particular value of
the disorder degree $\Delta/|J|=0.1$ this energy is small compared to the
characteristic energy scale in the figure ($\delta E_{c} \approx 0.001$).
The log-linear insert presents a blow-up of the distribution function
$P_{21}(\delta E)$ in the region of very small energy spacings
$\delta E \sim \delta E_{c}$ (each point is averaged over the interval
$\delta = 0.0005$ in the insert).

The quantity of interest in the present case is the first momentum
of the distribution $P_{21}(\delta E)$. Denote it as $E^*_2 -
E^*_1$.  The physical meaning of this quantity is clear: it is the
mean energy spacing between the levels of a local manifold. The
calculated disorder scaling of $E^*_2 - E^*_1 = \int d(\delta E)
P_{21}(\delta E) \delta E$ (diamonds) is shown in
Fig.~\ref{fig4}({\it b}) together with the best power-law fit
(dashed line) to the numerical data:
    \begin{equation}
        E^*_2 - E^*_1  \approx 0.40\; |J|\,\left(\Delta
        \over|J|\right)^{1.36}\ .
        \label{E21num}
    \end{equation}
Note that the disorder scaling laws for  $E^*_2 - E^*_1$ and
$\sigma_{11}$ are very close to each other. The theoretical
estimate of the $E^*_2 - E^*_1$ scaling was obtained in
Refs.~\onlinecite{Malyshev91,Malyshev95}:
    \begin{equation}
        E^*_2 - E^*_1  \approx (3\pi^2)^{-1/3}\;|J|\, \left({\Delta \over
        |J|}\right)^{4/3} \approx 0.32\;|J| \, \left({\Delta \over
        |J|}\right)^{4/3} \ ,
        \label{E21theor}
    \end{equation}
and agrees excellently  with the calculated behavior. This result was also
obtained numerically in Ref.~\onlinecite{Bakalis99}, where the detuning
between bleaching and induced absorption of the pump-probe spectrum was
simulated. This detuning is a good measure for the exciton delocalization
length~\cite{Malyshev95,Bakalis99}.

It is remarkable that rescaling of the numerical data for $E^*_2 -
E^*_1$ in terms of $f^*$ gives $E^*_2 - E^*_1 \sim (1/f^*)^2$, the
proportionality that holds for an ideal linear chain of length
$f^*$. This is yet another confirmation of the existence of the
local discrete energy structure of a weakly localized Frenkel
exciton near the lower band edge.

It is also of interest to compare the mean $E_2^* - E_1^*$ of the
$P_{21}(\delta E)$ distribution with the standard deviation
$\sigma_{11}$ of the $P_{11}(\delta E)$ distribution.
The figures~\ref{fig3}({\it c}) and \ref{fig4}({\it b}) as well as
the equations~(\ref{sigma}) and (\ref{E21theor}) for these
quantities show that $E_2^* - E_1^* < \sigma_{11}$. This
finding clarifies the fact that the local discrete energy structure can
not be seen neither in the density of states (see
Refs.~\onlinecite{Fidder93a,Fidder91,Schreiber82,Kozlov98}) nor in the
linear absorption spectra (see
Refs.~\onlinecite{Fidder93a,Malyshev99,Knapp84,Fidder91}); the energy
fluctuations of the local ground states exceed the characteristic
energy scale of the local energy structure, hence, the fine local
structure is hidden by the inhomogeneous  broadening of the local
ground states.

\begin{figure}[h!]
\includegraphics[height=2.5in]{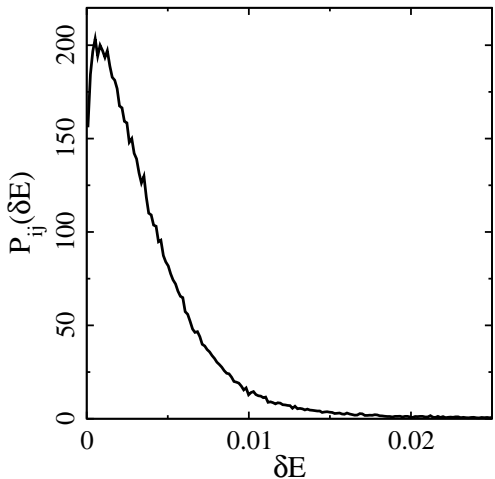}
\caption{
The distribution of the energy spacing between levels in the vicinity of the
exciton band bottom, $P_{ij}(\delta E)$, obtained for a 400 sites chain and
the disorder degree $\Delta/|J|=0.1$ (5000 realizations of disorder was
considered, each point of the distributions is averaged over the interval
(accumulation length) $\delta = 0.00015$). All energies are in units of
$|J|$. Note that that this distribution shows almost no level repulsion.
}
\label{fig5}
\end{figure}

The effect of level repulsion is well known from the studies of
spectra of complex systems, such as nuclei.~\cite{Mehta90} These
studies are often based on the random matrix theory
(RMT).~\cite{Mehta90} In particular, this theory predicts level
repulsion for the eigenvalues of a real symmetric matrix chosen at
random from the Gaussian orthogonal ensemble, i.e. when all matrix
elements are Gaussian stochastic variables statistically
independent of each other. Despite a seeming similarity of this
result and ours they cannot be directly related to each other.
First, the off-diagonal matrix element of the
Hamiltonian~(\ref{H}) are not stochastic variables while they are
in the random matrix theory. Moreover, in the RMT the off-diagonal
elements have zero mean, while in the Hamiltonian~(\ref{H}) they
are all negative. Second, we found the well-pronounced effect of
level repulsion only for a sub-ensemble of levels. Simulations for
the entire set of levels show almost no level repulsion in the
vicinity of the band bottom (see Fig.~\ref{fig5}).

\subsection{Long-range coupling effects}
\label{LR}

Complete set of the above mentioned simulations was also performed
for the exact DD coupling model.  It was found that $f^*$,
$\sigma_{11}$, and $E^*_2 - E^*_1$ are strongly affected by the
long-range dipole-dipole terms. The following expressions give the
best fit to the numerical data:
    \begin{mathletters}
    \label{3}
    \begin{equation}
        f^*  \approx 9.74\,\left({|J| \over \Delta}\right)^{0.80}\ ,
        \label{f*LR}
    \end{equation}
    \begin{equation}
    \sigma_{11} \approx 0.50\;|J|\,\left({|J| \over \Delta}\right)^{1.36}\ ,
    \end{equation}
    \begin{equation}
        E^*_2 - E^*_1  \approx 0.37\;|J|\,\left({\Delta \over
        |J|}\right)^{1.53}. \label{E21LR}
    \end{equation}
    \end{mathletters}
Unlike the NN model case, the rescaling formula $E^*_2 - E^*_1
\sim (1/f^*)^2$ does not hold for the exact DD model. This finding
is not surprising, as it agrees with the change in the exciton
energy spectrum of a regular chain in the DD approximation as
compared to that obtained in the NN approximation (see
Eqs.~(\ref{BottomLR}) and (\ref{BottomNN})). The logarithmic
factor in Eq.~(\ref{BottomLR}) is breaking the $E^*_2 - E^*_1 \sim
(1/f^*)^2$ dependence. The deviation of the exponents $0.80$ and
$1.53$ in Eqs.~(\ref{f*LR}) and~(\ref{E21LR}) from the NN model
values $2/3$ and $4/3$ is also due to this
factor.~\cite{Malyshev95} As a matter of fact, the power-law
approximation is not adequate for the exact model. Using the exact
low-energy spectrum of the 1D localized exciton  and the rule
described at the end of Sec.~\ref{gs_statistics}), the authors of
Ref.~\onlinecite{Malyshev95} obtained the correct equation for the
localization length $N^*$:
    \begin{equation}
        \frac {N^{*\,3/2}}{\ln N^* + a} = b\, \frac{|J|}{\Delta}
        \label{N*new}
    \end{equation}
with $a=3/2-ln(2^{4/3}\pi) \approx -0.57$ and  $b=3\pi^2 \approx 30$.

Straightforward comparison of the scaling laws of the quantities $f^*$ and
$N^*$ (the latter is defined as the number of sites under the main peak of
the wavefunction, see also the end of Sec.~\ref{gs_statistics}), shows that,
like in the NN model, these quantities are proportional to each other
$f^*=c\,N^*$, where $c\approx 0.62$. Fitting Eq.~(\ref{N*new}) to the
numerical data for $c^{-1}f^*$, the following optimal values of the
parameters were found: $a=-0.54$ and $b=25.1$. These values are in excellent
agreement with the theory.

In conclusion to this section we would like to comment on the
disorder scaling law of the exciton-radiative-rate enhancement
factor which was obtained numerically in
Ref.~\onlinecite{Fidder91} for the exact dipole-dipole coupling
model. This factor is introduced as follows: define the average
oscillator strength per state at energy $E$: $\mu^2(E) =
A(E)/\rho(E)$, where $A(E)$ and $\rho(E)$ are the absorption
spectrum and the density of exciton states respectively, then $max
[\mu^2(E)]$ is an effective measure for the enhancement of the
radiative rate. This quantity also contains information about the
spatial extension of the exciton states coupled to the
light.~\cite{Fidder91} The authors of Ref.~\onlinecite{Fidder91}
found from their simulations that $max [\mu^2(E)]\sim
(|J|/\Delta)^{0.74}$. The exponent of this scaling law, $0.74$,
differs from the value obtained in our simulations, $0.80$
(Eq.~(\ref{f*LR})). We believe that this discrepancy of the order
of 10\% originates in the different definitions of $f^*$ used in
our study and in Ref.~\onlinecite{Fidder91}.

\section{Summary and concluding remarks}
\label{Sum}

In this paper we discuss the statistics of energy levels of the 1D
Frenkel Hamiltonian with on-site randomness near the lower band
edge. The heuristic arguments on the existence of the local
discrete energy structure, which were put forward in
Refs.~\onlinecite{Malyshev91} and~\onlinecite{Malyshev95} are
confirmed by the detailed statistical study. Selecting the states
by means of the overlap integrals we find that the energy levels
of the well overlapping states indeed undergo repulsion resulting
in the local energy structure which is similar to the structure of
an ideal linear chain of the reduced length $N^*$ (localization
length). The lowest state of each set has no nodes within the
localization segment and therefore can be interpreted as the local
ground state, while the next state in a set has a node within the
segment and is the first local excited state, etc. The average
energy spacing of the local ground and first excited states
obtained within the framework of the NN model follows the
$(1/N^*)^{2}$ dependence that holds for an ideal chain of length
$N^*$.

On the contrary, the energy spacing of the non-overlapping local ground
states is characterized by zero mean and distributed in the energy interval
that is of the same order of magnitude as the mean energy spacing of the
local energy structure. Therefore, the local energy structure appears to be
hidden in the density of states~\cite{Fidder93a,Fidder91,Schreiber82} and in
linear absorption spectra as well.~\cite{Fidder93a,Knapp84,%
Fidder91,Schreiber82} Nevertheless, this structure can be revealed by the
pump-probe spectroscopy sharing two-exciton
states.~\cite{Malyshev95,Bakalis99} Indeed, due to fermionic nature of 1D
Frenkel excitons,~\cite{Chesnut63,Avetisyan85,Juzeliunas88,Spano91}
one-to-two and zero-to-one exciton optical transitions involving the states
of the same local structure would be blue-shifted with respect to each
other. This shift is equal to the energy spacing between the local ground
and first excited states. The blue shift of one-to-two and zero-to-one
exciton optical transitions in J-aggregates of pseudoisocyanine bromide was
observed experimentally~\cite{Minoshima94,Fidder93b} providing unambiguous
confirmation of the existence of the local discrete energy structure in the
vicinity of the exciton band minimum.

\acknowledgments

We are indebted to our relative Konstantin Malyshev who joined us in
performing this study and untimely passed away when this work was being
carried out. A. V. M. is grateful to ASCOL de Salamanca for support and la
Universidad de Salamanca for hospitality and computer facilities.

\appendix

\section{Effect of the cut-off parameters}

\begin{figure}[h!]
\includegraphics[height=2.75in]{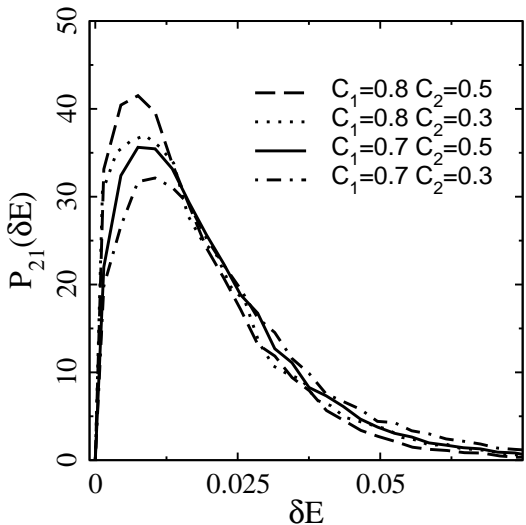}
\caption{
The distribution of the energy spacing between the local ground and first
excited states of the same segment, $P_{21}(\delta E)$, obtained for a 200
sites chain and the disorder degree $\Delta/|J|=0.1$ (40000 realizations of
disorder was considered, each point of the distributions is averaged over
the interval $\delta = 0.0025$). The curves are calculated for different
sets of the cut-off parameters:
$C_1=0.7$, $C_2=0.3$ --- dash-dotted line, 
$C_1=0.7$, $C_2=0.5$ --- solid line, 
$C_1=0.8$, $C_2=0.3$ --- dotted line, 
$C_1=0.8$, $C_2=0.5$ --- dashed line.
All energies are in units of $|J|$.
}
\label{fig6}
\end{figure}

In this section the weak dependence of the results on the cut-off parameters
$C_1$ and $C_2$ is illustrated. Figure~\ref{fig6} shows several
distributions $P_{21}(\delta E)$ calculated for different sets of the
cut-off parameters: $C_1=0.7$, $C_2=0.3$ --- dash-dotted line, $C_1=0.7$,
$C_2=0.5$ --- solid line, $C_1=0.8$, $C_2=0.3$ --- dotted line, $C_1=0.8$,
$C_2=0.5$ --- dashed line. These distributions are very close to each other,
which results in only weak (less than 15\%) quantitative dependence of
distribution momenta on the values of the cut-off parameters.

\pagebreak

\end{document}